\documentclass[]{aa}
\usepackage{graphicx}
\usepackage{natbib}
\usepackage{tabularx}
\usepackage{amsmath}

\nonstopmode
\begin{document}

\title{The curious case of the  companion:  evidence for  cold accretion onto a dwarf satellite near the isolated elliptical NGC 7796
 \thanks{Based on observations collected at the European Organisation for Astronomical Research in the Southern Hemisphere under ESO programme 60.A-9316(A).
    }}

\subtitle{}

\author{
T. Richtler   \inst{1}   
 \and 
M. Hilker \inst{2}
\and 
K. Voggel \inst{2,5}
\and
T.~H. Puzia \inst{3}
\and
R.Salinas \inst{4}
\and
M. G\'omez \inst{6}
\and
R. Lane \inst{3}
}

\offprints{T. Richtler}

\institute{
Departamento de Astronom\'{\i}a,
Universidad de Concepci\'on,
Concepci\'on, Chile;
tom@astro-udec.cl
\and
European Southern Observatory,
Karl-Schwarzschild-Str.2,
85748 Garching,
Germany
\and
Instituto de Astrof\'{i}sica,
Pontificia Universidad Cat\'olica de Chile,
Av.~Vicu\~na Mackenna 4860,
7820436 Macul, Santiago, Chile
\and
Gemini Observatory, Casilla 603, La Serena, Chile
\and
Department of Physics \& Astronomy,
University of Utah,
115 South 1400 East,
Salt Lake City,
UT 84112,
USA
\and
 Departamento de Ciencias F\'{\i}sicas, Facultad de Ciencias Exactas, Universidad Andres Bello, Fern\'andez Concha 700, Las Condes, CP 7591538, Chile
}

\date{Received  / Accepted }

\abstract
{
The isolated elliptical (IE) NGC 7796 is accompanied by an interesting early-type dwarf  galaxy, named NGC7796-DW1. It exhibits a tidal tail, very boxy isophotes, and multiple nuclei or regions (A, B, and C) that
 are bluer than the bulk population of the galaxy, indicating a younger age. These properties are suggestive of  
 a dwarf-dwarf merger remnant. 
   }
{
  We want to investigate the properties of the dwarf galaxy and its 
components to find more evidence for a dwarf-dwarf merger or for alternative  formation scenarios. 
   }
{We use the Multi-Unit Spectroscopic Explorer (MUSE)  at the VLT   to investigate 
NGC 7796-DW1.  We extract characteristic spectra to which we
apply the STARLIGHT population synthesis software to obtain ages and metallicities of the various population components of the galaxy.
    }
{ The galaxy's main body is old and metal-poor.  A surprising result is the extended  line emission in the galaxy, forming a ring-like structure with a projected diameter of 2.2 kpc.
The line ratios fall into the regime of HII-regions, although
 OB-stellar populations cannot be
 identified by spectral signatures.  The low H$\alpha$ surface brightnesses indicate unresolved star-forming substructures, which means that broad-band colours are not reliable
 age or metallicity indicators.
Nucleus A  is a relatively old (7 Gyr or older)  and metal-poor super star cluster,
 most probably the nucleus of the dwarf, now displaced. The star-forming regions B and  C show younger  and distinctly more metal-rich components. 
The emission line ratios of regions B and C indicate an almost solar  oxygen abundance, if compared with radiation models of HII regions. Oxygen abundances from empirical calibrations point
to only half-solar.
The ring-like H$\alpha$-structure does not exhibit signs of rotation or orbital movements. 
      }
{  
The dwarf-dwarf merger scenario is excluded because of the missing metal-rich merger component. A viable alternative is gas accretion 
from a reservoir of cold, metal-rich gas. NGC7796 has to provide this  gas  within its X-ray bright halo.  
Cold accretion may be a general
solution to the problem of extended star formation histories in transition dwarf galaxies.
 }
\titlerunning{dwarf companion}
\maketitle
\section{Introduction}

The first catalogue of isolated elliptical galaxies (IEs)
\citep{karach73} was based on photographic data and simply 
contained early-type galaxies with no apparent bright galaxy nearby (see  \citealt{richtler15} for a compact introduction to the literature).  IEs  in fact may have many dwarf  companions \citep{madore04}.  

\begin{figure}[]
\begin{center}
\includegraphics[width=0.51\textwidth]{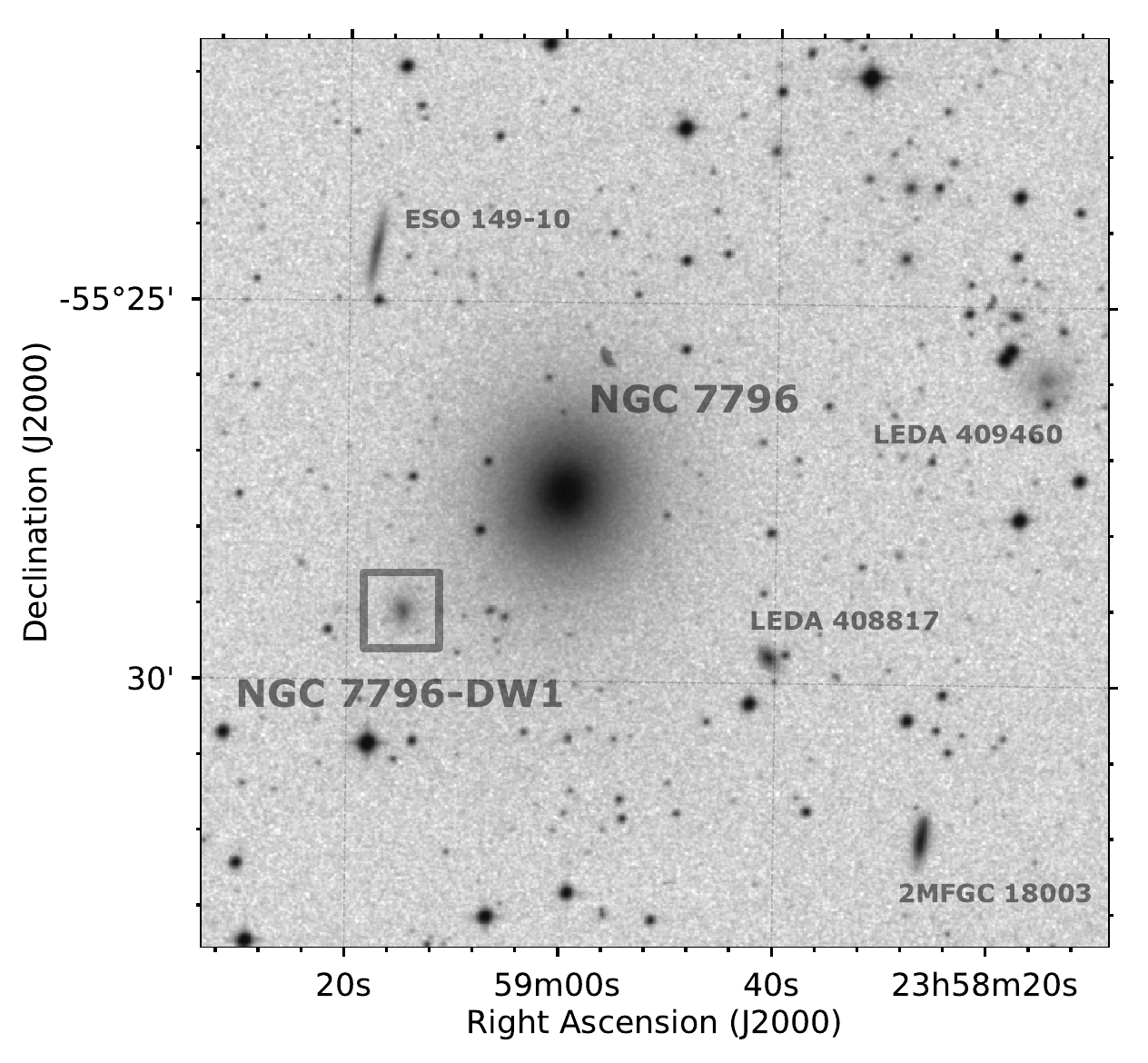}
\caption{The IE NGC 7796 and its dwarf companion, NGC 7796-DW1.}
\label{fig:fov}
\end{center}
\end{figure}

NGC 7796 is an IE \citep{reda04}. With respect to richness of the globular cluster system, photometric properties, and X-ray properties, it resembles a normal,
old elliptical galaxy in a group or cluster \citep{richtler15}, in contrast to many other IEs that show signatures of recent interactions and/or younger stellar
populations \citep{salinas15}. \citet{vader94} list two companions, but only NGC7796-1 (we  respect its nature as a dwarf galaxy and hereafter use  NGC7796-DW1)  can be confirmed as a real companion galaxy (see the remarks below) and is
classified as dE2.
 Figure \ref{fig:fov} shows NGC 7796 and its companion NGC7796-DW1 at a projected distance of about 2.1\arcmin or 30 kpc.   

A detailed image is shown in Fig. \ref{fig:linemorph} (upper left panel). The dwarf galaxy
 has  remarkable properties: it shows  a tidal tail (not visible here, but see \citealt{richtler15}),  and three  central sources which we interpret as  "multiple
nuclei" and label A, B, and C. However, only object A seems to be a point source, while objects B and C  are slightly extended. 
 Therefore only object A seems to be a real nucleus, and  we shall use  the more appropriate term "region" for the other two objects. 
The colours of regions B and C indicate intermediate ages of about
1 Gyr, whilst nucleus A and the main body of the galaxy appear older, but with ages that are difficult to estimate from only broad-band colours (more information in Section \ref{sec:populations}).
 The galaxy also  exhibits very boxy outer isophotes.  
 
These properties may characterise a merger remnant. Recent work points to the importance of dwarf-dwarf interactions or even dwarf-dwarf mergers:  the newly discovered  faint dwarf galaxy population in the Fornax cluster core
shows signs of correlated clustering \citep{munoz15,ordenes18}. Star formation
in pairs of dwarf galaxies is enhanced with respect to dwarfs  in isolation \citep{stierwalt15,pearson16}.
Moreover, shells in dwarf galaxies  in the Virgo cluster are best understood by dwarf-dwarf
mergers \citep{paudel17}.  
These examples show that dwarf-dwarf interaction/mergers are potentially important for the evolution of dwarf galaxies.

 NGC7796-DW1 appears as an example of a   transition dwarf galaxy  with a mixture of   late-type and early-type appearance (e.g. \citealt{sandage91,knezek99,lisker06,dellenbusch07,dellenbusch08,koleva13,
koleva14}).  

Here we present  a new study with  ESO's  new Multi-Unit Spectroscopic Explorer (MUSE),   revealing interesting results. The character of NGC7796-DW1 turns out to be very different from previous understanding. Detailed study of  ages, star formation history, and metallicities of its stellar population and the
surprising emission lines leads us to conclude that this dwarf galaxy has obtained its younger stellar populations not by a merger, but by accretion of gas suitable for star formation. 

 As in \citet{richtler15}, we adopt a distance of 50 Mpc based on the
%
surface brightness fluctuation  distance of \citet{tonry01}, corresponding to 242.4
pc per arcsecond.

\section{Observations and reductions}

NGC7796-DW1 has been observed with MUSE under the programme 60.A-9316(A), mounted at the Very Large Telescope (VLT) of the European
Southern Observatory (ESO) at Cerro Paranal, Chile.  The data were first taken as Science Verification data during the night 29/30 June 2014.   It turned out that with respect to
seeing and photometric quality they did not meet the required standards. The seeing was about 1\farcs5  and the night was not photometric. Therefore the observations were repeated during the night 19/20 August 2014, this time reaching a better seeing quality, but still not a high  photometric quality. For the present study, we only used the August data.
 
Table \ref{tab:observations} lists object, ESO Phase-3 designation, position, exposure times, and image quality as the FWHM-value of the  cube which we finally use.

MUSE is a mosaic of 24 Integral Field Units (IFUs), covering a field of 1$\times$1 arcmin$^2$ in the wide field mode. The pixel scale is 0.2$\times$0.2 arcsec$^2$. 
 The spectral  resolution varies from R=2000 at 4700 \AA\ to R=4000 at 9300 \AA\ which are the extremes of the spectral range covered.

\begin{table*}[ht!]
\centering
\begin{tabular}{cccccc}\hline
\hline
Object & Phase 3 data product & {Center Position RA} & {Center Position Dec} & {Exp.~Time(sec)} &{Seeing}\\
\hline
NGC 7796-DW1 & ADP.2017-03-23T15\_19\_01.004.fits & 23:59:14.73  (J2000)  & -55:29:04.7 (J2000) & 6$\times$900 s &  1\farcs 0  \\
\hline
\hline

\end{tabular}
\caption[Summary of MUSE observations]{{Summary of observations (ESO
    program ID 60.A-9316(A)). The f centre coordinates refer to the centre of an elliptical isophote with a radius of  5\arcsec measured on the indicated data cube.
    }}

\label{tab:observations}
\end{table*}

\subsection{Reduction}
The ESO Phase 3 concept offers through the ESO science archive reduced data products that  in our case are four MUSE observing blocks with  exposure times of 2$\times$900 s each. 
  Originally, we tried to work with a  data cube that uses all exposures, so we performed an independent reduction. 
However, we found that the inclusion of the June data does not  improve the final results,    
so for reasons of clarity and reproducibility, we used the Phase 3 final data cube which only contains the data from August 2014. This data product with the
designation given in Table \ref{tab:observations}  is available in the ESO science archive. For completeness, we briefly describe the reduction process.
 The reduction has been performed using the pipeline provided by ESO, described by the pipeline manual version 1.6.2, employing the standard EsoRex recipes. 
 The basic reduction consists of applying {\it muse\_bias} and {\it  muse\_flat} (no correction for dark currents), followed by the wavelength-calibration with {\it  muse\_wavecal}. 
 The line-spread-function has been calculated from the arc spectra using {\it muse\_lsf}. For the instrument geometry, we used the tables provided by ESO. Twilight exposures
 were used for the illumination correction, applying {\it muse\_twilight}.   The previous recipes produce frames/tables that  now enter the recipe {\it muse\_scibasic} that performs bias subtraction,
 flat field correction, wavelength calibration,  and more. The recipe {\it muse\_scipost} performs flux calibration and calculates the final data cube or as a choice, fully reduced pixtables that are combined by {\it muse\_exp\_combine} to
 produce a data cube with combined individual exposures. The pipeline corrects also for  telluric absorption features.
     
 Sky subtraction in the Phase 3 pipeline used the sky exposures and left a weak residuum of the light of NGC7796 itself and strong sky lines. To remove the sky almost completely, we defined a local sky in the
 field free of other sources
 and performed the sky subtraction for each spectrum individually using {\it skytweak} under
IRAF.  Our sky spectrum has been extracted at the position  RA=23:59:15.946, Dec.=-55:29:31.0 with
 QFitsView and a radius of 60 pixels.

 The data are not designed for absolute photometry.
  Since only one standard star is used in the
 pipeline, we do not expect it to be   very precise. We will not draw strong conclusions from absolute fluxes.
 We expect the relative flux uncertainties to be approximately or better than  5\% which is the overall experience (e.g. \citealt{weilbach15}).
We use QFitsView \citep{ott12}  for extracting and averaging spectra. 
The uncertainties for each wavelength element have been calculated using the STAT part of the MUSE  cube which gives the variance for each pixel and wavelength element.

\subsection{Astrometry and coordinate correction}
The absolute astrometric accuracy (for identification purposes only) of the MUSE
World Coordinate System (WCS) is relatively good. The 
USNO star
U0300.38162244 (RA: 23:59:13.374, Dec: -55:29:20.87) can be found in the MUSE field. To agree with its coordinates, the MUSE WCS was shifted 1.5\arcsec\ to the north,  0.11s 
  to the east.
These are the coordinates which we use.

The literature coordinates  of the two dwarf companions of NGC7796 are probably incorrect. \citet{vader94} give coordinates for J1950, which precessed
for J2000 would indicate that NGC7796-DW1 is not identical with APMUKS  B235639.76-554544.9 but with a stellar-like object displaced by about 20\arcsec\
to the south-west (it is indeed a star). The coordinates for NGC7796-2 indicate a faint galaxy, but shifted by the same displacement as for NGC7796-DW1,  and point to
the clear case of a background spiral. We thus conclude that the coordinates of \citet{vader94} (which appear in the NED) must be corrected by 11\arcsec\ to the east and
16\arcsec\ to the north.

\section{Line emission and its morphology}


\begin{figure}[]
\begin{center}
\includegraphics[width=0.51\textwidth]{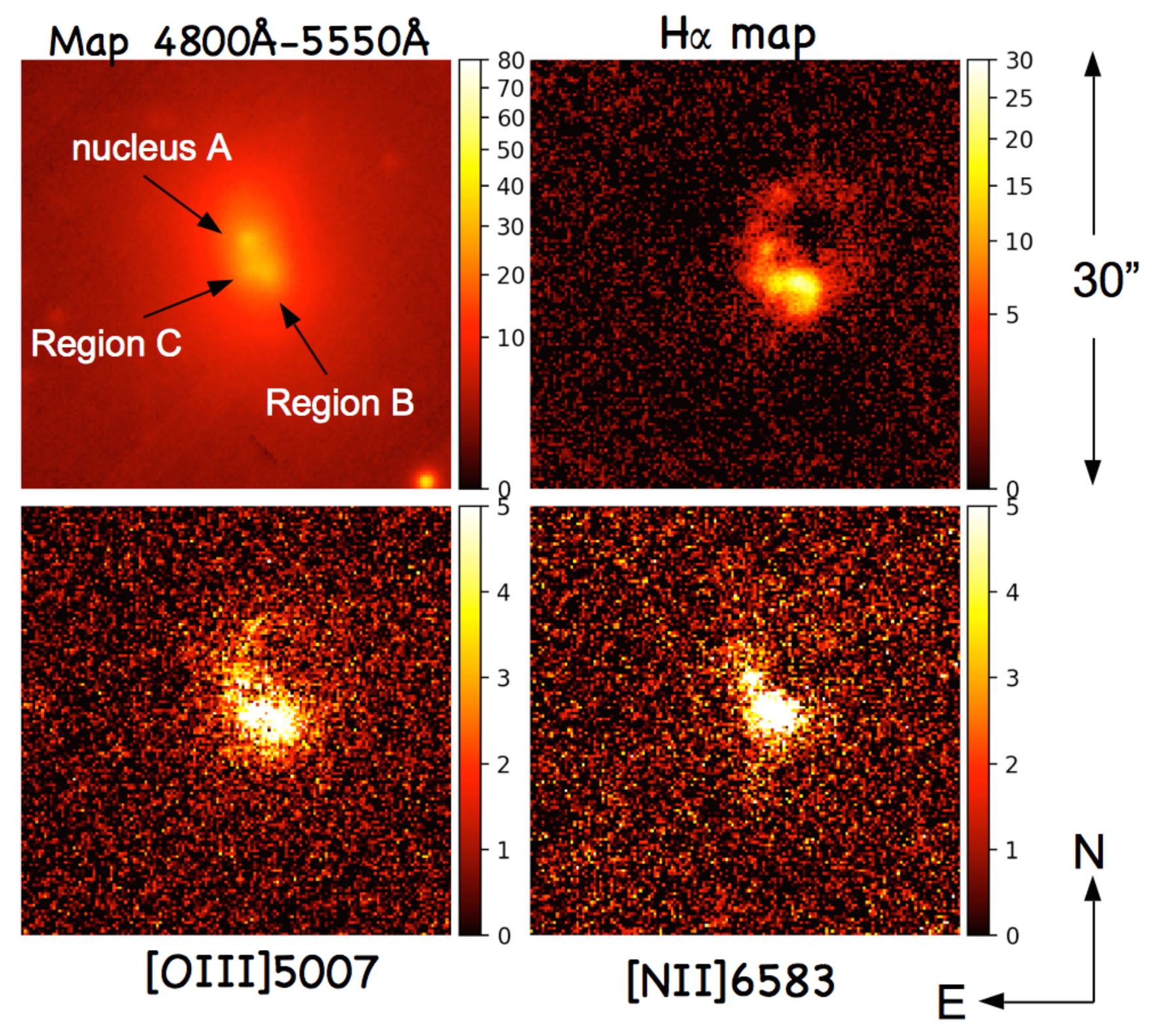}
\caption{ Upper left panel: a  MUSE image of NGC7796-DW1 showing the three different nuclear regions with a square root scaling. Indicated is the mean flux in the interval 4800\AA-5550\AA. As in the other
maps, the unit is $10^{-20} erg/s/cm^2/$ \AA. Upper right panel: a map of the  H$\alpha$-flux.  The H$\alpha$-emission defines   an almost  complete,
but distorted ring with some swellings and knots.
 Lower left panel: [OIII]5007 map. The  [OIII]5007/ H$\alpha$  ratio is constant which indicates
local ionising sources. Lower right panel: The [NII]6583-map.}
\label{fig:linemorph}
\end{center}
\end{figure}

The field around NGC7796-DW1 covered by MUSE,  showing nucleus A and the regions B and C, is displayed in the upper-left panel of Fig. \ref{fig:linemorph}.
 On the VIMOS B-image with its better seeing of 0.8\arcsec\  \citep{richtler15}, region B appears extended, but the true size is difficult to measure. We estimate a FWHM of about 1\arcsec, and 
 it is clearly visible that only nucleus A can be a real nucleus or super star cluster. In an outer isophote with a half major axis of 8\arcsec\ and a  half minor axis  of 5.5\arcsec\, nucleus A is offset from the centre by about 1\arcsec\ (or 240 pc) along
the major axis to the north.

A  surprising observation is the presence of extended line emission in a major  region
of the visible galaxy. 
The other panels of Fig. \ref{fig:linemorph} show  maps of  H$\alpha$, the [OIII]5007\AA-line, and the [NII]6583\AA-line. 
The H$\alpha$-emission is the brightest, which is already an indication that the ionisation resembles more an HII-region than 
LINER-like line ratios. These maps have been produced by integrating over the full line widths using the line map option of QFitsView with the local continuum subtracted.
The H$\alpha$ emission forms a complete, somewhat  elongated ring with the brightest parts in region B. The ring  is not complete in the other line maps. The diameter is about 9\arcsec or 2.2 kpc.
 The ring shows inhomogeneities or swellings, best visible in H$\alpha$ where  continuum-structure-like point sources are not visible. The H$\alpha$ intensity may therefore reflect the gas density rather than the ionising flux.  The emitted flux from nucleus A is not higher than the local background, so nucleus A itself is not a source of line emission as regions B and C are. Its position is also offset from  the ring. Further remarks on the ring are given in section \ref{sec:ring}.

Besides nucleus A and the regions B and C,
we want to analyse a spectrum from outside these regions, but near the centre where the signal-to-noise ratio (S/N) is still high. Therefore we extracted a spectrum with  the
coordinate from Table \ref{tab:observations} as the centre and a radius of 15 pixels (or 0.727 kpc), and subtracted the spectra of nucleus A, region B, and region C, named R15-minusABC. 
 Furthermore,
 we investigate a  spectrum of the outer parts of the galaxy,  which is a spectrum of the annulus 15-30 pixels, named R30-R15. 
 Finally, we analysed a spectrum  covering the entire galaxy with a radius of 40 pixels, named R40 (or 1.939 kpc).
  These six  spectra are described in Section \ref{sec:populations}.

\section{Population synthesis}

\label{sec:populations}
For the spectral analysis, which ideally  gives ages and metallicities of the population components and, moreover, produces a model spectrum that can be subtracted for the emission line analysis, 
we employ the {\it STARLIGHT} code (\citealt{cid05,mateus06}, Version 4). This code fits flux-calibrated spectra using linear combinations of library model spectra. The outputs  are mass- and luminosity weighted 
 components, characterised by age and metallicity.  Reddening, radial velocity,  and fit-related parameters are made available as well. The code has been widely applied in the CALIFA survey
 \citep{cid11,kehrig12, gomes16,zinchenko16}.

 STARLIGHT does not provide uncertainties. The population decomposition of S/N-limited spectra  is plagued by  degeneracies between age, metallicity, reddening, spectral resolution, spectral range, and so on.
The MUSE spectra are quite red and less appropriate for investigating young populations. 
An important point
 is that the broad-band  colours of the younger objects in our dwarf do not straightforwardly  indicate their ages.    A seeing of 0.8\arcsec\  may mix unresolved young blue structures 
with the galaxy light and thus  let the colours appear redder than the bluest subcomponents. This effect is even more pronounced in the case of the MUSE spectra with their seeing of about 1\arcsec. Therefore the task at hand is to answer the question of  how much of the young components are visible in the spectra, rather than an analysis of the true star formation history.

\begin{figure*}[th!]
\begin{center}
\includegraphics[width=1.0\textwidth]{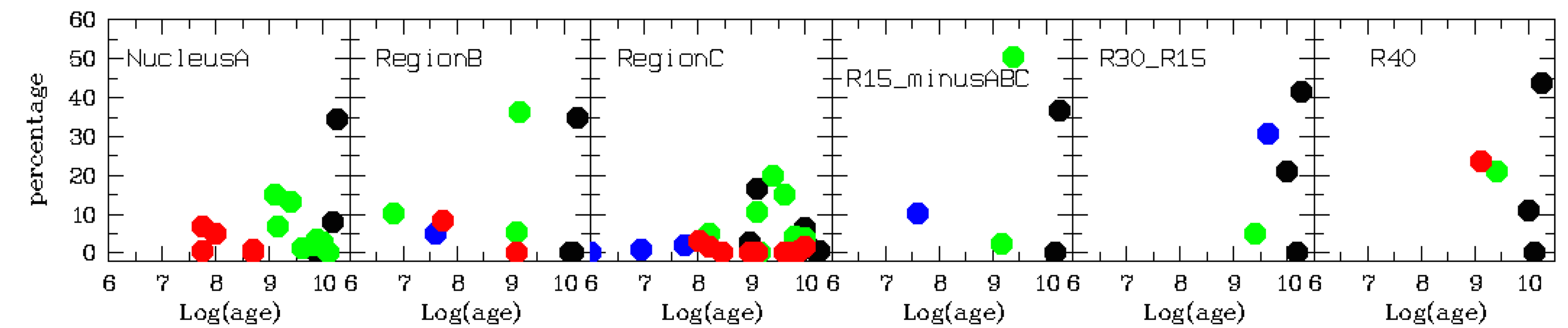}
\caption{These graphs show typical results of a STARLIGHT decomposition for our characteristic spectra. Indicated is the
population composition in percentage of light at  5650 \AA\  of all populations involved in the fit. The colours denote metallicities. 
Black: z=0.0004; blue: z=0.008, green: z=0.02, red: z=0.05.  The individual uncertainties, particularly those at low percentage values, are probably high
and determined by many systematics. 
Therefore this must not be confused with a determination of the star formation history. Rather it confirms what has been suggested by broad-band
photometry.  Further discussion of this can be found in the text. }
\label{fig:starlight}
\end{center}
\end{figure*}

Our spectral library consists of 125 model spectra from \citet{bruzual03} with solar-scaled composition. The metallicities range from z=0.0004 to z=0.05, and the ages range from 0.001 Gyr to 18 Gyr. 
We do not iterate to reduce the number of statistically significant components by deleting all components below a given percentage limit, but show all components
that have been involved in the fit. 
The strong emission lines (H$\beta$, H$\alpha$, [OIII] 4959\AA\ and 5007\AA , [NII] 6549\AA\ and  6583\AA , [SII] 6717\AA\ and  6731\AA)  are masked out. The properties of the population synthesis modelling of each co-added spectrum are listed in  
Table \ref{tab:fits} and displayed in Fig. \ref{fig:starlight}.
  Table \ref{tab:fits} lists the identification, the extraction radii, the absorption in the V-band, the radial velocity, $\chi^2$, and two S/N values provided by STARLIGHT. The lower values have been determined
  from a direct comparison with the model spectra and thus may be the most trustworthy. The higher values result from using the variances  of each pixel which are given in the STAT part of the MUSE cube. 
  For each spectrum, we extracted the corresponding spaxels in the STAT section and calculate the  resulting variance as $\sigma^2$ = N$\times \sigma_i^2$, where N is the number of extracted spaxels and
  $\sigma_i^2$ the individual variances. The final variance after sky subtraction is the sum of object+sky and sky variance.   The square roots of these variances 
  are used by STARLIGHT in the fit. 

\begin{table*}
\begin{tabular}{ccccccccc}
\hline
\hline

ID & Name & R/B-R  & radius (pixels) &  $\chi^2/N$ & SN1/SN2  &norm.factor[10$^2$]  & $V_r$[km/s] & $\sigma$[km/s]     \\     
\hline
1  &  nucleus A &  21.1/1.06  &    2   &  0.68 & 18/22 &  3.09 &  3621.6  &   1.0        \\  
2  & region B  &    21.3/0.65 &    3         &   0.81 &   22/29 & 5.95 & 3622.3 &  0  \\
3  & region C &   21.5/0.83   &     2        &   0.71  &  19/18  &  2.52 & 3644.2 & 18  \\
4  & R15minusABC &  & R15-ABC    &  0.39 &  23/32 &  95.9 & 3623.2& 15 \\  
5  & R30-R15 &  - & 15-30  & 2.52  & 18/104 &  179 & 3618.9 & 7  \\
6  & R40       & -  & 40  & 2.27  & 22/115  & 340 & 3622.2  & 10  \\
\hline
\end{tabular}
\caption{Table lists properties and fit results of our characteristic spectra. Name, R-magnitude colour B-R, extraction radius, $\chi^2/N$. SN1 is the S/N that STARLIGHT finds from comparison with the
model spectra, and SN2 is the S/N that corresponds to the flux uncertainty provided by the STAT part of the MUSE cube. The normalisation factor is used for calculating absolute fluxes.
The last columns are the heliocentric radial velocity and the velocity dispersion.  Magnitude and colour are from \citet{richtler15}. We give the kinematical data without interpretation.} 
\label{tab:fits}
\end{table*}

\subsection{Nucleus A}
The spectrum of nucleus A   has been extracted using the circular option of  QFitsView with a radius of 2 pixels. 
 The spectrum  
 is dominated by a bright stellar continuum with a strong absorption of H$\beta$ (see Fig.\ref{fig:nucleusA}), where 
  the emission line is hardly visible.
The emission lines vanish almost completely if a spectrum of the local background is subtracted, which shows that the object itself does not exhibit line emission.
Figure \ref{fig:starlight}  indicates a moderately metal-poor and intermediate-age to old object.     The weak young contributions  may therefore arise from background/foreground populations. 
If we select only components older than 1 Gyr from Fig. \ref{fig:starlight}   the weighted mean values for age and metallicity  are  8.7 Gyr and z=0.004, respectively.  
With $M_R = -12.4,$ according to \citet{richtler15}, the associated mass is about several $10^6 M_\odot$, characterising an $\omega$ Centauri-like object \citep{hilker00}. In analogy, one would expect a range of populations to be present.  
Although NGC7796-DW1 may possess a  globular cluster system on its own, we cannot clearly identify  individual point sources as globular clusters, but
an object as massive as  nucleus A cannot be a normal cluster. It is much more plausible that it was the (now decentred) nucleus of the dwarf elliptical (see more remarks in section
\ref{sec:transition}).
\subsubsection{Ca-triplet}
 It is of interest to try an independent metallicity estimate using the calcium triplet at 8498\AA,\, 8542\AA,\, and 8662\AA.  In this near-infrared region, contributions from younger populations are negligible.
The calibration of \citet{sakari16} indicates that for stellar populations older than 2 Gyr, the sum of the three equivalent widths (EWs) is independent  from age and only sensitive to metallicity. Moreover,
\citet{dacosta16} emphasises the very weak dependence on [Ca/Fe].  The main uncertainty is the definition of the continuum.    Unfortunately, the redshift of NGC7796  superimposes the  lines 8542\AA\, and 8662\AA\ onto
strong sky features. Figure \ref{fig:nucleusA} shows that the continuum of the best-fitting population is placed somewhat too high
and the lines themselves are not well fitted. The higher spectral resolution of MUSE in the infrared
may also be responsible for the bad representation. We fit the continuum `manually' with {\it splot} in IRAF  between 8470\AA\ and 8780\AA\ and estimate the following EWs: EW(8498)=1.4\AA,  EW(8542) = 2.2\AA, 
EW(8662)=2.2\AA.   Unresolved spectral features may somewhat increase the EWs and we estimate the uncertainty in all lines to be about 0.2\AA. 
The calibration of \citet{sakari16} is
\begin{equation}
\label{eq:catcal}
[Fe/H] = 0.38(\pm0.1) \times EW(\Sigma CaT) -3.48(\pm0.13)
.\end{equation}
This results in a  metallicity of  [Fe/H]= $-$1.3$\pm$0.6 dex, where also the uncertainties of the coefficients have been considered.  This is obviously only a weak constraint, the dominant error source
being the slope coefficient in Eq. \ref{eq:catcal}. 

\subsection{Region B}
\label{sec:regionB}
This region is the brightest spot in H$\alpha$, but, as Fig. \ref{fig:linemorph} shows, the intensity peak of the pure emission is somewhat displaced from the optical image. 
Its spectrum is displayed in Fig. \ref{fig:regionB}. The population synthesis yields metal-poor  old components, but  metal-rich and very young components as well.  The   metal-rich intermediate-age contribution is much stronger than in the case of nucleus A. 

There are plausibly several populations that are mixed together by the seeing. The emission line spectrum is that of a typical HII region, so OB populations must be present.  The detection of dust would be
interesting.
The Balmer decrement indicates a weak increase from the standard value of 2.85, but because the reddening law is unknown, the reddening itself remains uncertain.  The values in Table \ref{tab:fluxes} 
have been calculated by  applying the standard relation 
from \citet{momcheva13}: $ E(B-V) = 1.97\times log_{10}(H_\alpha/H_\beta)/2.85) $.

\subsection{Region C}
The broad-band colour of region C is somewhat redder than that of region B (Table \ref{tab:fits}). Whether it  is older cannot be decided from our data.   The very young components are  indeed missing. Metal-poor, intermediate-age components are dominating.
 The strong emission lines have an
appearance  very similar to those of region B.  The H$\alpha$ peak luminosity is one third of that of region B.
 We conclude that also region C is actively star forming, but at a lower rate.


\subsection{R15minusABC}
We want to analyse a spectrum that samples the inner part of the dwarf galaxy but largely avoids the younger regions. Therefore we extract a spectrum with a radius of 15 pixels and subtract the spectra of nucleus A,  and regions B and
C.  Figure \ref{fig:starlight} shows that now intermediate-age and old metal-poor components are dominant. Calculating the mean for all components older than 1 Gyr, one obtains an age of 9.1 Gyr and a metallicity of z=0.005.

  
\subsection{R30-R15}
This spectrum is the extraction of an annulus between of 15 and 30 pixels in radius. It samples the outer parts of the galaxy and the low-luminosity zone of the north-western parts of the
H$\alpha$ ring. Only old and metal-poor components are found. The mean values for age and metallicity are  11 Gyr and z=0.002. These values are indicative only. However, they convey the impression of
an old, moderately metal-poor dwarf galaxy.


 

\subsection{R40}
We extract the total spectrum with a radius of 40 pixels, encompassing an area of about  5000 spaxels.  Apart from a measurement of the total H$\alpha$-luminosity,
it is 
 a useful consistency check for comparison with the other spectra that are more specific regarding the composition of  stellar populations.
Indeed, no  young population contributes, but STARLIGHT finds a range of intermediate-age populations of moderate metal deficiency.






\section{Emission lines and diagnostic graphs}

\subsection{Line fluxes and diagnostic graphs}

\label{sec:emission}

 The lower panels of the spectra depicted in the appendix show the emission line spectra after the model spectra have been subtracted. In Fig.\ref{fig:nucleusA} the emission lines
are identified. One finds the usual strong lines. Atomic neutral emission lines are not visible.

We  measure the line fluxes in the model-subtracted spectra using {\it splot} in IRAF between those wavelengths where the wings of the emission lines reach zero.
 The fluxes are measured in the spectra normalised to the model spectra.  Absolute flux values are then calculated by  applying a factor given by STARLIGHT. These factors are given in Table \ref{tab:fits}.

The line fluxes and  line ratios that we use in strong line diagnostics are  given in   Table \ref{tab:fluxes}.  
 The Balmer decrements may indicate the existence of dust in star-forming regions. 
Balmer decrements H$\alpha$/H$\beta$ range from 2.6 (spectrum 5) to 3.2 (spectrum 4). 
The reddening values have been calculated using a  formula from 
\citet{momcheva13} (see section \ref{sec:regionB}).
The outer regions  should  be dust free, therefore this is qualitatively expected. It is encouraging that the Balmer decrement in the brightest sources, regions B and C, is very close to the
standard value of 2.85. This shows a precise separation of stellar and emission line parts in the spectra.  

 Comparison with photoionisation
models that predict strong line fluxes depending on the ionisation rate q\footnote{The ionisation rate has the unit number of ionising photons per unit area per second and per particle density.}
and oxygen abundance shows good agreement
 \cite{dopita13} give a summary of the history of strong line diagnostics  and  a series of diagnostic graphs of which only a few are
interesting for us because of the restricted wavelength range. They recommend in particular the graphs [OIII]/H$\beta$ versus [NII/][SII] and [OIII]/[SII] versus [NII/[SII], which permit
the cleanest separation between  ionisation rate and  oxygen abundance. Here, the measurements refer to the sum of both [SII] lines and the [NII] line  at 6584\AA , respectively.
In addition we show the graph  [OIII]/H$\beta$ versus [NII]/$H\alpha$ to demonstrate that the line ratios remain within the range expected for HII regions.

\begin{figure*}[]
\begin{center}
\includegraphics[width=0.9\textwidth]{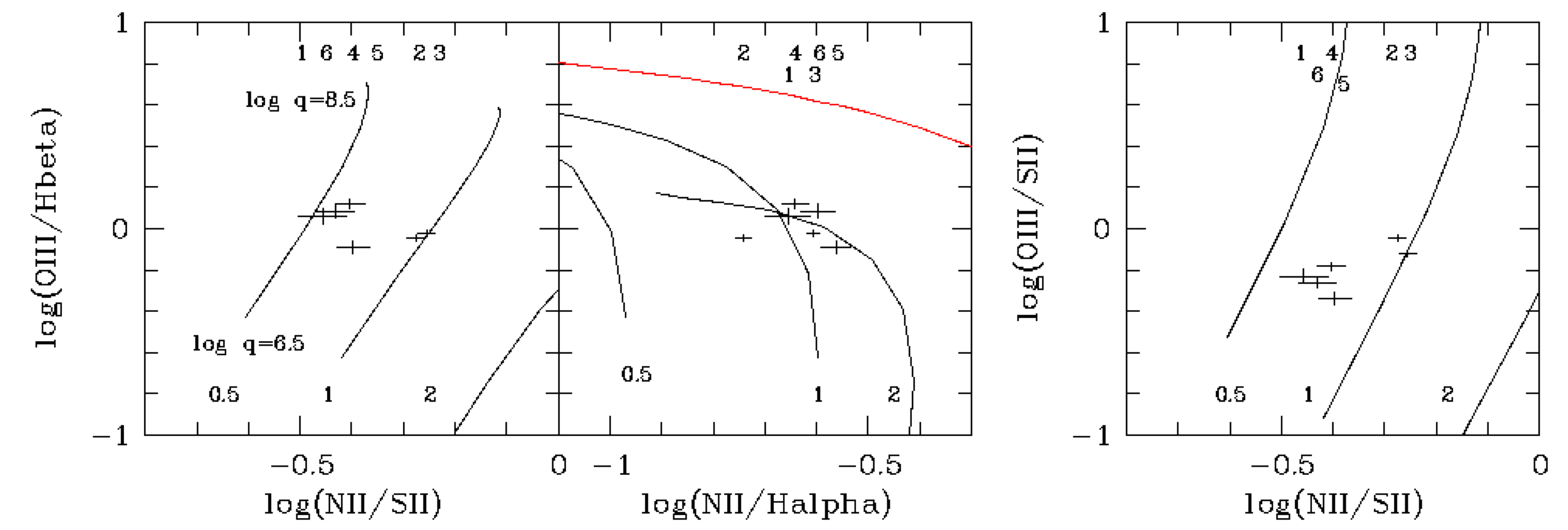} 
\caption{{  Diagnostic graphs for our six characteristic  spectra. These line ratios
 are compared
to HII-region models from \citet{dopita13} (for a Boltzmann distribution of electrons) with the ionisation parameter log (q) and  three oxygen abundances in solar units (0.5, 1, 2)  as parameters. The solid lines are lines of
constant abundance and varying log (q). The range of
log (q) values is indicated in the left panel and is  the same for all panels.   The solid red line in the middle panel is the line
separating HII-regions from AGN-like spectra according to 
\citet{kewley01}.  The numbers in the upper parts identify the spectra from Table \ref{tab:fluxes} by their sequence. The symbol sizes represent the measurement uncertainty.
 As  noted by  \citet{dopita13}, the right diagnostic graph seems to provide the cleanest separation 
of log (q) and z.  
The spectra indicate a solar oxygen abundance  for regions B and C and slightly less for the other spectra  (Table \ref{tab:abundances}).
This abundance variation is not, however, backed up 
by the empirical calibration of \citet{pilyugin16}.
} }
\label{fig:diagnostics}
\end{center}
\end{figure*}

A new feature in the models of \citet{dopita13}  is that the electron energies follow a $\kappa$-distribution which in the case of $\kappa=\infty$ is equal to a Boltzmann distribution. 
According to \citet{ferland16}, the thermalisation of electrons with a finite $\kappa$-value occurs  on shorter timescales than heating or cooling which makes $\kappa=\infty$ the preferred choice.
Figure \ref{fig:diagnostics} shows  HII-region models of three different oxygen abundances (0.5, 1, 2; in solar units) for a Boltzmann distribution of electron
energies. The ionisation rates  range from log(q)=6.5 to log(q)=8.5.  A good separation of log(q) and oxygen abundance z can be seen in the left and right panels.
The middle graph  is less suitable for separating these
parameters. The  solid red line  marks the   limits between the HII-regions and active galactic nucleus(AGN)-dominated ionisation from \citet{kewley01}.  It is now widely acknowledged that  post-AGB stars 
from older populations also provide an ionising radiation field (e.g. \citealt{cid11}) that is not distinguishable from LINERS  produced by AGNs. The  distinction line is 
convincingly explained  by the fact that high metallicities result in lower temperatures of HII regions and therefore show a degeneracy in the zone of high abundance.
\citet{zhang17}
 present photoionisation models of intermediate-age and old stellar populations for some strong-line diagnostic diagrams. Their diagrams also indicate that the ionising sources in our case  are OB stars.
 
For interpolating the tables of \citet{dopita13}, we use linear expressions in the intervals of interest, which are log(q) $<$ 8.0 and  0.5$<$z$<$3.   For the left panel, we obtain
\begin{equation}
\begin{split}
z = 1.7 (\pm0.04) +2.85(\pm0.08) \times \log{ [NII]/[SII]} \\
 -0.85(\pm0.03) \times \log{ [OIII]}
\end{split}
,\end{equation}
and for the right panel, we obtain 
\begin{equation}
\begin{split}
z = 1.8 (\pm0.06) + 3.2(\pm0.1) \times \log{ [NII]/[SII]} \\
 -0.6(\pm0.03) \times \log{ [OIII]/[SII]}
\end{split}
.\end{equation}
 We do not give values for the middle panel.  

The resulting oxygen abundances  are listed in Table \ref{tab:abundances} 
in the notation 12+log(O/H) with the solar abundance given by 12 + log(O/H)=8.69 \citep{grevesse10}.
 The uncertainties are the
standard deviations calculated from error propagation of the uncertainties in Table \ref{tab:fluxes}. These small errors
do not respect systematics. The uncertainty in the reconstruction of H$\beta$  is probably  an important systematic  effect.
However, one observes only a small scatter in [OIII]/H$\beta$ and moreover, the abundances are much more affected
by a shift in the other indices. 
The abundance difference of about 0.25 dex between spectra 2 \&\ 3 and 4 \&\ 5 is striking.   However,
this difference is not visible in other calibrations. 



The mean abundance of 0.6 solar is also consistent with the high metallicities of regions B and C found in the
population synthesis. 



\begin{table*}
\resizebox{0.9\linewidth}{!}{
\begin{tabular}{cccccccccc}
\hline
\hline
 ID & spectrum  & H$\beta$   & [OIII]  &H$\alpha$ &[NII] &    [S II] &  [S II] & E(B-V)  & log(SB(H$\alpha$)) \\
 & wavelength &    4861      & 5007  & 6563      & 6582 & 6715 & 6731   &    &  erg/s/kpc$^2$   \\
\hline
1 & nucleus A   &  2.32$\pm$0.15    &  2.66$\pm$0.15    &  7.14$\pm$0.15   & 1.58$\pm$0.15    &  2.63$\pm$0.15  &  1.89$\pm$0.15   & 0.066$\pm$0.008  &    -  \\
2 & region B    &   17.52$\pm$0.30   &  15.9$\pm$0.3  & 52.0$\pm$0.30 &  9.42$\pm$0.3 &   10.8$\pm$0.3  &  6.97$\pm$0.3  &   0.034$\pm$0.002   & 38.33\\
3 & region C   &   6.48$\pm$0.13 &      6.2$\pm$0.13  &    18.2$\pm$0.13    & 4.52$\pm$0.13  &   4.84$\pm$0.13   & 3.30$\pm$0.13   & 0   & 38.23 \\
4 & R15minusABC & 106$\pm$5 &  140$\pm$5 &          370$\pm$5 &          84$\pm$5   &   124$\pm$5 &   90$\pm$5   &  0.17$\pm$0.005   & 37.84\\
5 & R30-R15 &  202.0$\pm$9.3 & 164$\pm$9 &  525$\pm$9      & 144$\pm$9 &  206$\pm$9  & 153$\pm$9  & 0   & 37.48 \\
 6 & R40  &  291.5$\pm$17.5 &  353$\pm$17 &  947$\pm$17.3     & 239$\pm$17.3 & 385$\pm$17 & 260$\pm$17.3 & 0.1$\pm$0.007   & 37.36  \\   
\hline
\end{tabular} 
}
\caption{Absolute emission line fluxes that have been measured in the spectra normalised by STARLIGHT.  The unit is $10^{-18} erg/cm^2/s$. The relative uncertainties have been calculated by estimating the uniform   
uncertainty of 0.05 for the normalised spectra and multiplied by the normalisation factors which are given in the last column.
The absolute uncertainties are unknown and may suffer from mediocre photometric quality.  The reddening has been calculated from the Balmer decrement using the relation of \citet{momcheva13}.}
\label{tab:fluxes}
\end{table*}



\subsection{Empirical calibration}
\citet{pilyugin16} provide   
an empirical calibration using  three strong line indices  that reproduce the oxygen abundances  of a sample of calibrating HII regions with  temperature-based
abundances that have a very  small scatter  of  the order 0.01 dex.
They calibrate the following indices: N2 = $(6548+6583)/$H$\beta$, S2 = $(6715+6731)/$H$\beta$, R3 = $(4958+5007)/$H$\beta$, where the wavelengths stand for the fluxes of the respective lines (they calibrate further indices involving [O{\sc ii}] 3727\AA\ that are not relevant here). 
Because   we can measure H$\alpha$ more precisely than we can H$\beta$, we adopt H$\alpha\!=\!2.89\times$H$\beta$ for N2 and S2, to be consistent with \citet{pilyugin16}.   The resulting oxygen abundances from equation (6) of \citet{pilyugin16} (upper branch)  for our spectra are listed in column 5 of Table \ref{tab:abundances}. The mean abundance is  12+log(O/H) = $8.36$  or
0.5 solar with an obviously very small scatter. 

This is encouraging, given the very inhomogeneous measurements of the line indices particularly of  H$\alpha$ which depends on the accuracy of
the fitting by STARLIGHT. 
It may  be that the relative oxygen abundance systematically differs from  the pattern of the calibrating HII regions. 
Finally, there is the long-standing problem that empirical calibrations give lower oxygen abundances than the photoionisation models (e.g. (\citealt{bresolin09}); we also refer to
the discussion section of \citet{pilyugin16}).
Again, it is safe to say that
the oxygen abundance compared with HII regions fits much better to regions B and C than to the other spectra.

\subsection{Nitrogen-to-oxygen-ratio}
The nitrogen-to-oxygen-ratio  is used in the discussion below of whether or not a high N/O ratio can fake a high oxygen abundance (see section.\ref{sec:transition}).  

We apply the calibration of \citet{zahid12} (their equation (3))  
\begin{equation}
\begin{split}
log(N/O) = -0.86 + 1.94\times log([NII]/([SII]_1 +[SII]_2) \\ + 0.55\times (log([NII]/([SII]_1 +[SII]_2))^2
\end{split}
.\end{equation}

The resulting values are listed as the last column in Table \ref{tab:abundances}.

\begin{table*}[h!t]
\resizebox{0.7\linewidth}{!}{
\begin{tabular}{cccccc}

\hline
\hline
 ID   &     spectrum & equation (1) & equation (2) & \citet{pilyugin16} &   log(N/H) \\
\hline
        1 & nucleus A &  8.23$\pm$0.17 & 8.37$\pm$0.14       &       8.25$\pm$0.04  &      -1.64  \\
        2  &  region B &  8.68$\pm$0.02  &  8.67$\pm$0.03     &      8.32$\pm$0.01    &     -1.33  \\
        3  & region C & 8.68$\pm$0.02   &   8.71$\pm$0.02   &       8.37$\pm$0.01   &     -1.38  \\
        4 & R15-minusABC & 8.34$\pm$0.08 & 8.48$\pm$0.06 &  8.34$\pm$0.03 &    - \\
        5  & R30-R15 & 8.50$\pm$0.06   &   8.55$\pm$0.06     &      8.30$\pm$0.03   &    -1.51  \\
        6  & R40  &  8.29$\pm$0.11    &    8.45$\pm$0.09    &     8.33$\pm$0.03   &    -1.66  \\
        
\hline
\end{tabular}
}
\caption{Oxygen abundances  from comparison of strong line ratios  with  models and from empirical calibrations.    Columns 3 and 4 list the oxygen abundances in solar units for the six spectra  derived from equations (1) and (2). Column 5 (unit: 12+log(O/H))  lists the
abundances from equation (6) of \citet{pilyugin16}. The nitrogen abundance
has been calculated using equation (3). }
\label{tab:abundances} 
\end{table*}

\subsection{H$\alpha$-luminosity, total stellar mass}


 With the adopted distance of 50 Mpc  and the H$\alpha$-flux from Table \ref{tab:fluxes}, the total  H$\alpha$-luminosity becomes  $2.69\times10^{38}$ erg/s  (before absorption). The uncertainty
 of this value is considerable, the actual  measurement error being only a minor source. 
 We do not know the
 reddening law, and simply adopting the mean Galactic value (identifying H$\alpha$ with the R-band), that is,  $A_R = 2.3\times E(B-V) = 0.23$ (e.g. \citealt{cardelli89}, see their Table 1 for the scatter of $R_V$-values),
 may be quite incorrect. The absorption factor in the R-band ranges from 1.2 to 1.5. Adopting a measurement uncertainty of 10\% for the absolute flux, the flux interval is (2.9-4.4)$\times10^{38}$ erg/s. 
 The distance uncertainty probably dominates the other sources.   If
 one optimistically adopts 10\% uncertainty for the distance then the fluxes corresponding to the distances 55 Mpc and 45 Mpc span the range (2.3-5.4)$\times10^{38}$ erg/s.

 Can this   H$\alpha$-luminosity  be produced by evolved stellar populations as ionising sources?  If all Lyman continuum photons from populations older than $10^8$ yr are absorbed and re-emitted in the hydrogen recombination lines, 
 \citet{cid11} calculate  the expected H$\alpha$-luminosity  (their Eqs. (2) and (3)) as
 
\begin{equation}
L_{H\alpha} =  \frac{E_{H\alpha}}{f_ {H\alpha}}  M_{stellar} q_H
\label{eq:expected}
.\end{equation}

In this equation, $E_{H\alpha}=3.026\times10^{-12}$ erg is the energy of an $H\alpha$ photon, $f_ {H\alpha} = 2.206$ determines the fraction of energy emitted in $H\alpha$, 
$M_{stellar}$
is the total initial mass, 
derived from STARLIGHT's parameter {\it mcortot}, and $q_H = 10^{41} /sec/M_\odot$ is the rate of hydrogen ionising photons. 
If we approximate  the total stellar mass using the \citet{marigo08}-models
from paper II with the population parameters   [Fe/H]=-0.7 and an age of 2.5 Gyr, the M/L-value in the R-band is 1.1 and the mass corresponding to $M_R = -17.8$ becomes $6.9\times10^8 M_\odot$.
The expected H$\alpha$-luminosity is subsequently 9.5$\times10^{37}$ erg/s, that is, much lower than the observed one. If we furthermore take into account that the  H$\alpha$-luminosity is actually produced in only a small fraction
of the galaxy's volume, our adopted stellar mass is much too high for a realistic estimation. One concludes that the ionising sources must be an OB-star population that is unresolved.

%




\subsubsection{HII regions and diffuse ionised gas}

It is interesting to compare the emission line properties of our spectra with those  of \citet{zhang17} who investigate a large sample of star-forming MaNGa galaxies with the intention of distinguishing
between effects of HII regions and diffuse ionised gas (DIG). The mix of DIG and emission lines from HII regions may influence diagnostic diagrams of strong lines and so bias, for example, a metallicity 
determination. The morphological main difference between DIG and HII regions in their sample is the H$\alpha$-surface brightness, a low surface brightness being characteristic of DIG.  Regarding line
ratios,  DIG  resembles  LINER-like line ratios. Comparing the H$\alpha$ surface-brightness values from Table \ref{tab:fluxes} with those of \citet{zhang17}, 
one notes that Table \ref{tab:fluxes} contains only values comparable to the lowest surface brightnesses in the MaNGa sample.   On the other hand, the line ratios are well covered by the
grids of  HII regions, while the DIG regions of \citet{zhang17} are not. 

That the H$\alpha$ surface brightnesses are so low may be well explained by spatially unresolved substructure.

If there was contamination by DIG towards high metallicities, we would expect it to be most important in spectra 4 and 5 that sample gas outside the peak luminosities of regions B and C.
But the metallicity of spectra 4 and 5 is lower than that of spectra 2 and 3, meaning that at least a shift to higher metallicities through the influence of DIG is not obvious. 

An interesting observation is that spectrum 5,  which in its emission part samples the fainter outer region of the H$\alpha$-ring,  also shows a log (q) that is relatively similar to that of regions B and C, although
the surface brightness is much lower. A direct conclusion is that the ionising sources, although fainter, must be local.  A comparison with Fig.20 of \citet{zhang17} shows that in the  
diagnostic diagram of log([OIII]/H$\beta$) versus log([SII]/H$\alpha$), both spectra 4 and 5  are still consistent with HII regions and not with  a very old population 
as the ionising source.

\subsection{Does the ring rotate?}
 \label{sec:ring} 
\begin{figure}[]
\begin{center}
\includegraphics[width=0.5\textwidth]{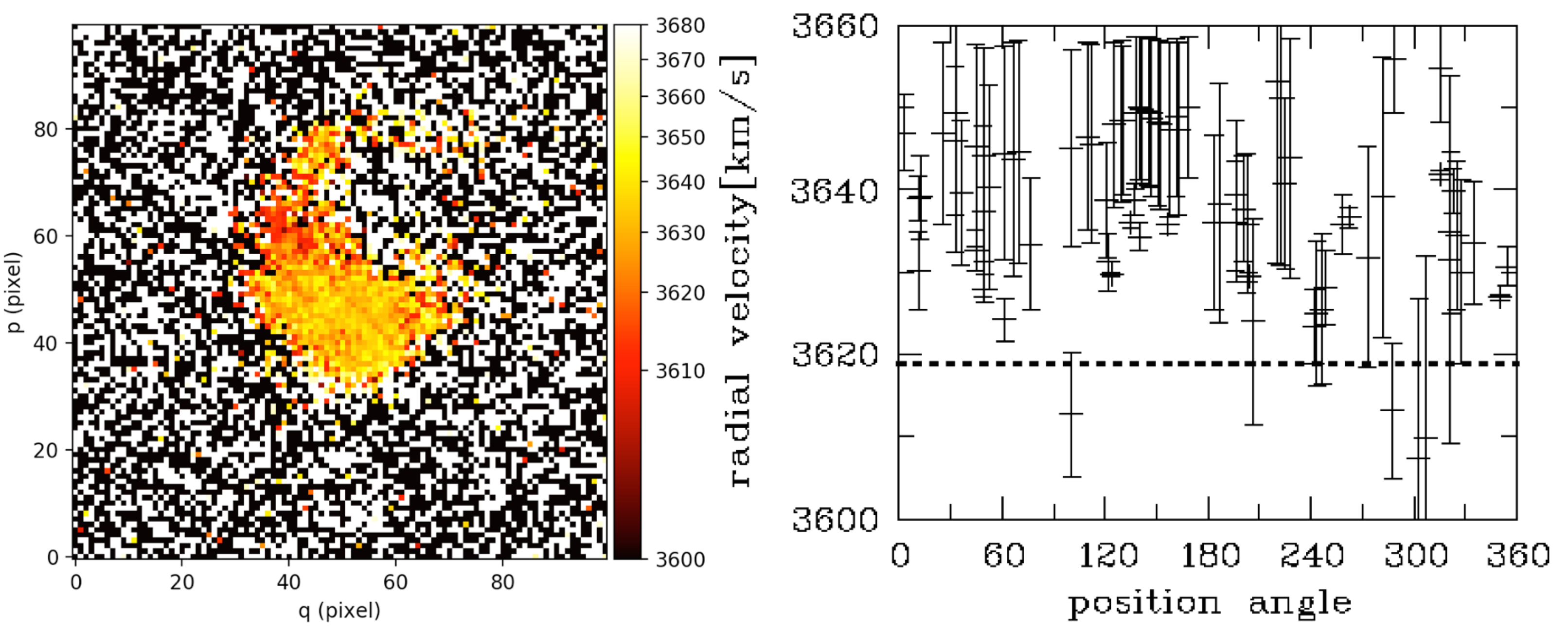}
\caption{Left panel: H$\alpha$ radial velocity map. Right panel:  radial $H\alpha$ velocities vs. position angle along the ring. These extractions overlap slightly and give velocities also in the regions of low surface brightness.
The horizontal dotted line is the stellar radial velocity of spectrum 5, assumed to be the systemic velocity.    A  rotation signal is
not apparent. }
\label{fig:ringkinematics}
\end{center}
\end{figure}

 A ring defines a plane in galaxies that is normally connected with orbital motion, for instance a rotating disk.  The natural question therefore arises of whether signs of rotation are visible. Due to the
 faintness of  emission lines in the north-eastern part, only H$\alpha$ provides the possibility to measure radial velocities along the ring structure. The left panel of Fig. \ref{fig:ringkinematics} 
 shows the H$\alpha$ radial velocity map.  The gas shows little velocity structure, and the low-surface-brightness zones are not easily visible.        
  In the right panel we plot 81 radial velocities versus three 
 position angles, defined east of north.       
 The spectra have been
 extracted with very different numbers of spaxels because of the very different surface brightnesses. The horizontal dotted line is the stellar velocity of spectrum 5, assumed to be the systemic velocity.
 It is apparent that the velocity is not constant, but a rotational velocity structure   should not have the same systemic velocity as the lower bound. The gas rather seems to move with a bulky substructure. 
Data of higher precision would be very useful in constraining the geometry of any such structure.

\section{Discussion}
\label{sec:discussion}

\subsection{The context of transition dwarf galaxies}
\label{sec:transition}
 
    NGC7796-DW1 is clearly an
 early-type dwarf galaxy  with actual or recent star formation, often labelled as a transition dwarf galaxy. The high metallicity of the interstellar medium might be a general indication for accreted material.
  Here we give a compact literature review. 
  We start with IC 225 for which  \citet{gu06} determined an oxygen abundance distinctly higher than that expected from the mass-metallicity relation.  \citet{dellenbusch07, dellenbusch08} presented a further six low-mass, oxygen-rich  early-type galaxies with recent star formation.  A major  element of
 their  discussion involved the question of where
 to locate these galaxies in an evolutionary sequence between dwarf irregulars and dwarf spheroidals. Influential work was presented by \citet{grebel03} who used the position in the luminosity-metallicity
 diagram as well as the HI mass
 to identify some 
 transition dwarf galaxies on their evolutionary paths towards dwarf spheroidal galaxies. 
 
 This scenario was supported by \citet{lisker06,lisker07} who found a subgroup 
 of disk-like objects among
Virgo dwarf ellipticals, sometimes with spiral structure,  whose spatial distribution in the cluster  resembles irregular and spiral galaxies. Because the cluster environment plausibly triggers  the 
 transition from late-type to early-type galaxies, one may conclude that also mass-poor galaxies can undergo transitional processes. 
 Subsequently \citet{peeples08} followed with a list of dwarf galaxies extracted from the catalogue of \citet{tremonti04} and apparently deviating from a general mass-metallicity relation.
 The high oxygen abundance has been interpreted by \citet{berg11} as an effect caused by a high N/O ratio. However, \citet{zahid12} correct for an enhanced N/O ratio and still find a high scatter
 in the oxygen abundance at low stellar mass, up to about solar metallicity. In our case, the nitrogen abundances from Table \ref{tab:fluxes} can be compared with those in Fig. 10 of \citet{zahid12}
 which shows that they are not enhanced.
 
 The transformation from early-type to late-type galaxies might also be possible. \citet{kannapan09} and \citet{wei10} identify E/S0 galaxies on the blue sequence and conclude that disk growth
 in early-type galaxies may occur (see the introduction of \citealt{wei10} for a compact review of cold gas in early-type galaxies).  
 
 \citet{hallenbeck12} searched for HI in early-type Virgo dwarfs, finding 12 objects with an HI content typical for dwarf irregulars. Among them, five objects show indications of star formation. They
 discuss several evolutionary processes for these galaxies and conclude that re-accretion of gas from cosmic filaments is the most plausible process.  



For our object, the merging scenario faces the serious difficulty that only one part of the imagined pre-merger galaxies can be identified. 
 Explaining the presence of young, metal-rich components requires a host galaxy of a high mass that corresponds to the higher metallicity. With the bulk of its stars having a metallicity of about [Fe/H]=$-$0.7, NGC7796-DW1 fits well to the mass-metallicity relation of dwarf galaxies \citep{smith08}, but the metal-rich stellar component is minor in comparison to the metal-poor component, while it should be  more massive. 

  Moreover, the time sequence of the star formation poses problems for the merger scenario. 
 Considering a merger with two components, a nucleated
dwarf elliptical and an irregular dwarf, the merger and coalescence should have happened less than 0.1  Gyr ago,  because in merger events the peak of star formation occurs well before
coalescence \citep{stierwalt15}.  Such a relatively young merger should not leave a dwarf with undisturbed isophotes and developed tidal tails.  This is a point also made by \citet{dellenbusch08}
in their morphological study of transition-type dwarfs.    
   
Therefore, excluding a merger,  only one viable scenario remains: accretion of cold gas from the inner gaseous halo of NGC 7796, 
a galaxy this is old  and X-ray bright. \citet{osullivan07} gives, using XMM,
an abundance profile that shows nearly solar abundance in the central region and then declines modestly, but with uncertainties which still  permit almost solar abundance at larger radii. Cold gas in early-type,
X-ray-bright  
galaxies   probably has cooled down from the hot phase \citep{werner14,smith17}. It  may
reside in disks \citep{serra16,young18},  but a disk in NGC7796 has not yet  been identified.

  Accretion of cold gas with subsequent star formation is a process whose importance for galaxy evolution has been long underestimated. In cosmology, simulations show the accretion of
pristine gas from filaments \citep{dekel09} which trigger star formation in high-z galaxies.  For local galaxies,   \citet{sancisi08} review the process of accretion of cold gas
for spiral galaxies with respect to morphological features, but do not mention dwarf galaxies.  However, "tadpole" galaxies nicely demonstrate the effect of cold-gas accretion on
the morphology of dwarf galaxies \citep{sanchez13}. Another strong indication for infall of cold gas is the existence of  metal-poor star-forming regions in otherwise moderately
metal-poor dwarf galaxies \citep{sanchez15}.





 

 


\subsection{Remarks on the star formation history of dwarf galaxies}
One classical early-type dwarf galaxy with an extended star formation history is the Andromeda companion NGC 205 with its young nucleus \citep{melcher87,dacosta88,monaco09}. 
Practically all early-type local-group dwarf galaxies exhibit extended
star formation histories (SFHs).  A complete literature compilation and extensive discussion  is found in  \citet{weisz14} who investigated
the SFHs in about 40 local-group dwarf galaxies.  An interesting trend (with exceptions) is
that the more massive dwarfs form a lower stellar mass fraction at early times than the less massive dwarfs, a  behaviour that is opposite to the "downsizing" of giant galaxies.
Weisz et al. explain that due to environmental effects, low mass dwarf spheroidals are preferably found within the virial radii of their host galaxies, while more massive dwarfs populate a
larger radius interval. Low-mass galaxies are therefore more susceptible to tidal effects and ram pressure stripping which may quench star formation relatively early.
 
However,   NGC7796-DW1 is a counterexample to this scenario by showing tidal effects and is also exposed to ram pressure from the hot X-ray gas of NGC 7796. This may therefore suggest that gas accretion  is the
main driver of star formation in this dwarf galaxy. A dependence on galaxy mass is qualitatively very plausible by assuming that accretion onto deeper potential wells works more efficiently and not only
in the central region where one expects the cold gas phase to be densest.

\section{Summary and conclusions}
We investigate the dwarf companion NGC 7796-DW1 of the IE NGC 7796 by means of IFU data obtained with MUSE at the VLT. The objective is to
further  probe a possible  dwarf-dwarf merger as suspected by \citet{richtler15} due to the presence of multiple nuclei and probably a multitude of
stellar populations. We distinguish the bulk of the dwarf galaxy: nucleus A and the bluer regions B and C.  

The first and surprising finding is that the galaxy, although classified as
early-type, is filled with
ionised gas, the emission lines being superimposed on top of a stellar continuum or in the case of Balmer lines, embedded in partly deep absorption troughs.  Best 
visible in H$\alpha$, the emission forms a ring with a diameter of 9\arcsec\ or 2.2 kpc of strongly varying brightness.

 We define a set of characteristic spectra by sampling the
"nuclei"  and the stellar population of the galaxy at different radii to achieve a possibly high S/N of the extracted spectra. 
We performed a population synthesis using STARLIGHT   \citep{cid05} that fits linear combinations of library spectra to  an observed spectrum. 

Nucleus A with a  mass of about $5\times10^6 M_\odot$ is the best candidate for the nucleus of this  early-type dwarf galaxy. The population synthesis identifies a strongly metal-poor, old component with an intermediate-age contribution.   
Region  B is the region with the highest H$\alpha$-luminosity.  Its stellar spectrum is still dominated by intermediate-age and old populations, but also shows young ($<$ 10$^8$ yr) and metal-rich contributions.  
It is not possible to assign a unique age, but the spectral appearance is the result of spatially unresolved subpopulations.
 Region C as the secondary peak in the H$\alpha$ luminosity has a similar metallicity, but the young components are weaker. 

 STARLIGHT indicates some absorption in regions B and C.     Possible
ongoing star formation is below visibility.  
However, a small reddening results in a better agreement between photometric and spectroscopic ages. 
Younger populations are still present outside the brightness peaks of regions B and C. The outer parts, and therefore the bulk of the galaxy, resemble the population of nucleus A in that they are metal-poor and are of old to intermediate age. 

The emission line strengths can be measured after subtracting the stellar model spectra. 
The line ratios are typical for metal-rich star-forming regions.  We compared the line strengths with HII models and empirical calibrations. The oxygen abundance is close to solar  which is consistent with the metallicity of the younger stellar populations in the galaxy.
OB stars are excluded as ionisation sources, leaving post-AGB stars as the most plausible alternative.  This shows that line-ratios might not be reliable 
 for distinguishing between ionising radiation fields of OB stars and post-AGB stars.  

The main argument against a dwarf-dwarf merger scenario is the population structure. The mass-metallicity relation suggests that regions B and C cannot have been part of a more metal-rich dwarf;  the dwarf must dominate by mass whereas the dominating component is more metal-poor than the comparably low-mass regions B and C.

Star formation in our dwarf galaxy must therefore have been triggered by infall of metal-rich gas. A plausible reservoir of gas with appropriately high
metallicity is the gas of the X-ray halo in NGC7796. We therefore conclude that star formation in our dwarf has been triggered by accretion of cold gas that
coexists with the hot X-ray gas.  The ring can be understood as the debris of a disk that formed from infalling gas with a certain angular momentum. We suggest that this might be a general recipe for complex SFHs of dwarf galaxies.

\begin{acknowledgements}
We thank the referee for his/her thorough report, in particular for pointing out  the misinterpretation of broad-band colours. 
TR acknowledges support from  CONICYT project  Basal AFB-170002. 
 He also acknowledges an ESO senior visitorship, during which the data reductions were performed,  which was essential for this contribution.
T.H.P. gratefully acknowledges support by FONDECYT Regular Project No. 1161817 and by CONICYT project Basal AFB-170002.
Dominik Bomans is thanked for discussions on STARLIGHT and literature hints.
We use QFitsView (version 3.2, developed by Thomas Ott and Alex Agudo Berbel at the Max Planck Institute for Extraterrestrial Physics, Garching)  for extracting and averaging spectra.
\end{acknowledgements}

\bibliographystyle{aa}
\bibliography{paper_dwarf.bib}

\begin{appendix}
\section{Sample spectra}

As sample spectra, we show only the spectra of nucleus A, region B, and R30-R15, whose most significant variation is seen in their population properties.

\begin{figure*}[]
\begin{center}
\includegraphics[width=0.9\textwidth]{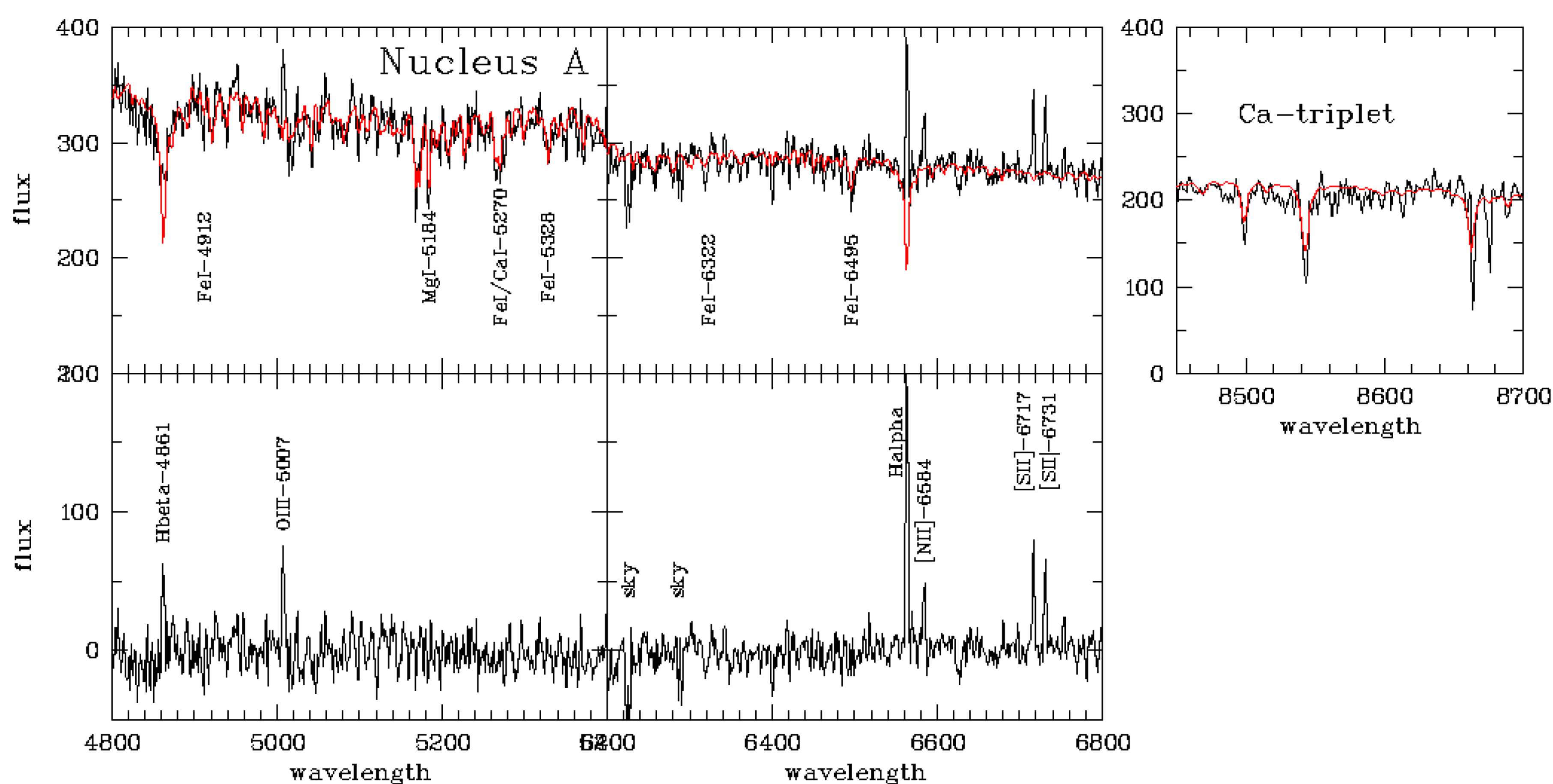}
\caption{The spectrum of nucleus A (see section \ref{sec:populations}). We show only the two  most interesting wavelength intervals. Some spectral features are marked for orientation.  Upper panel: the black solid line marks the observed spectrum, 
the red solid line the model spectrum. The fluxes  are given in units of $\mathrm 10^{-20} erg/cm^2/ \AA$. 
Lower Panel: The model spectrum has been subtracted. Here the emission
lines are local foreground/background and the cluster itself apparently does not host ionised gas.  Right panel: Spectral region of the Calcium triplet. This near-infrared region is not well fitted. We therefore
measure the equivalent widths directly in the spectrum.}
\end{center}
\label{fig:nucleusA}
\end{figure*}

\begin{figure*}[]
\begin{center}
\includegraphics[width=0.9
\textwidth]{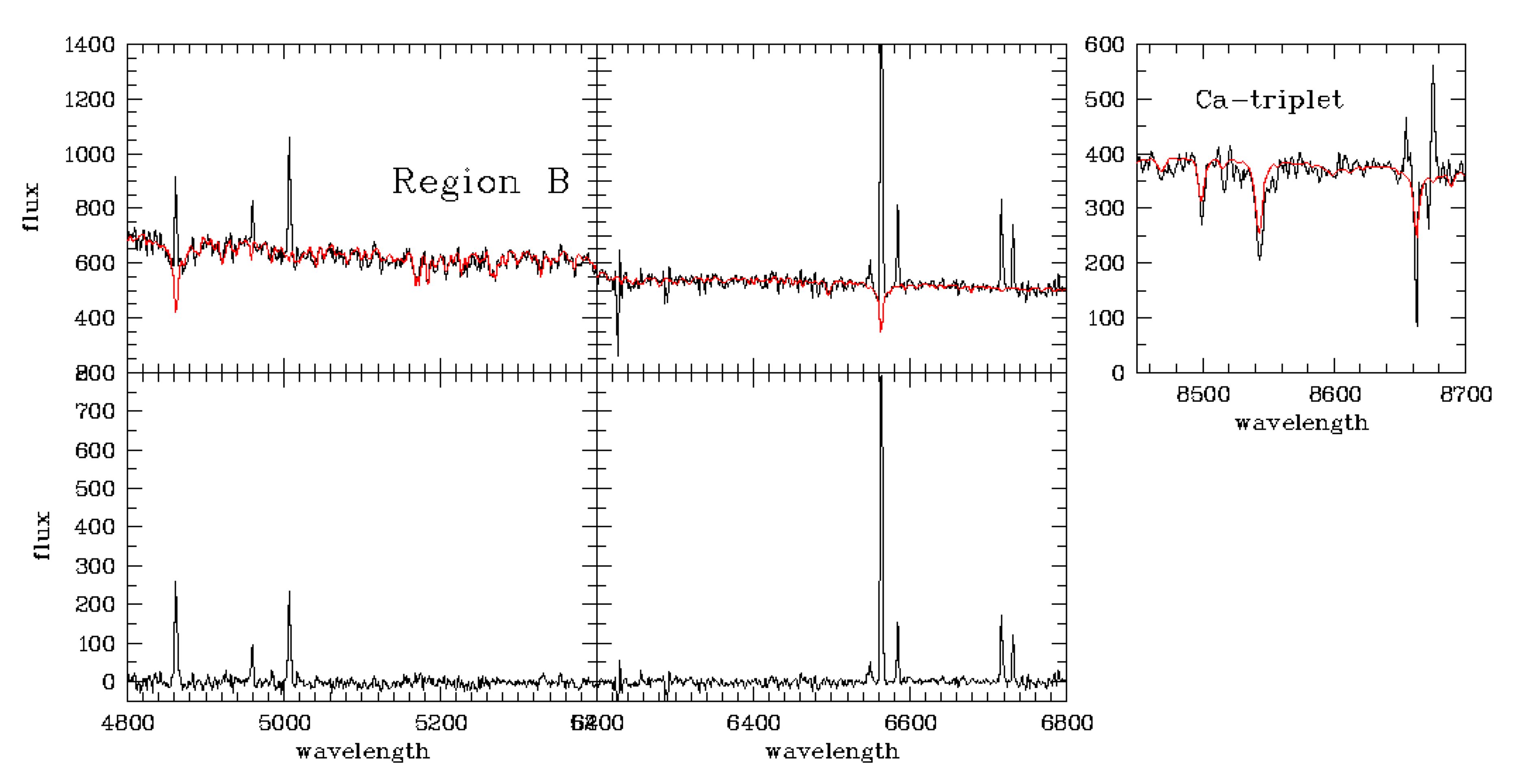}
\caption{The spectrum of region B.  This is the brightest region. We refer to the caption of Fig. \ref{fig:nucleusA} for more information.  }


\end{center}
\label{fig:regionB}
\end{figure*}

\begin{figure*}[]
\begin{center}
\includegraphics[width=0.9
\textwidth]{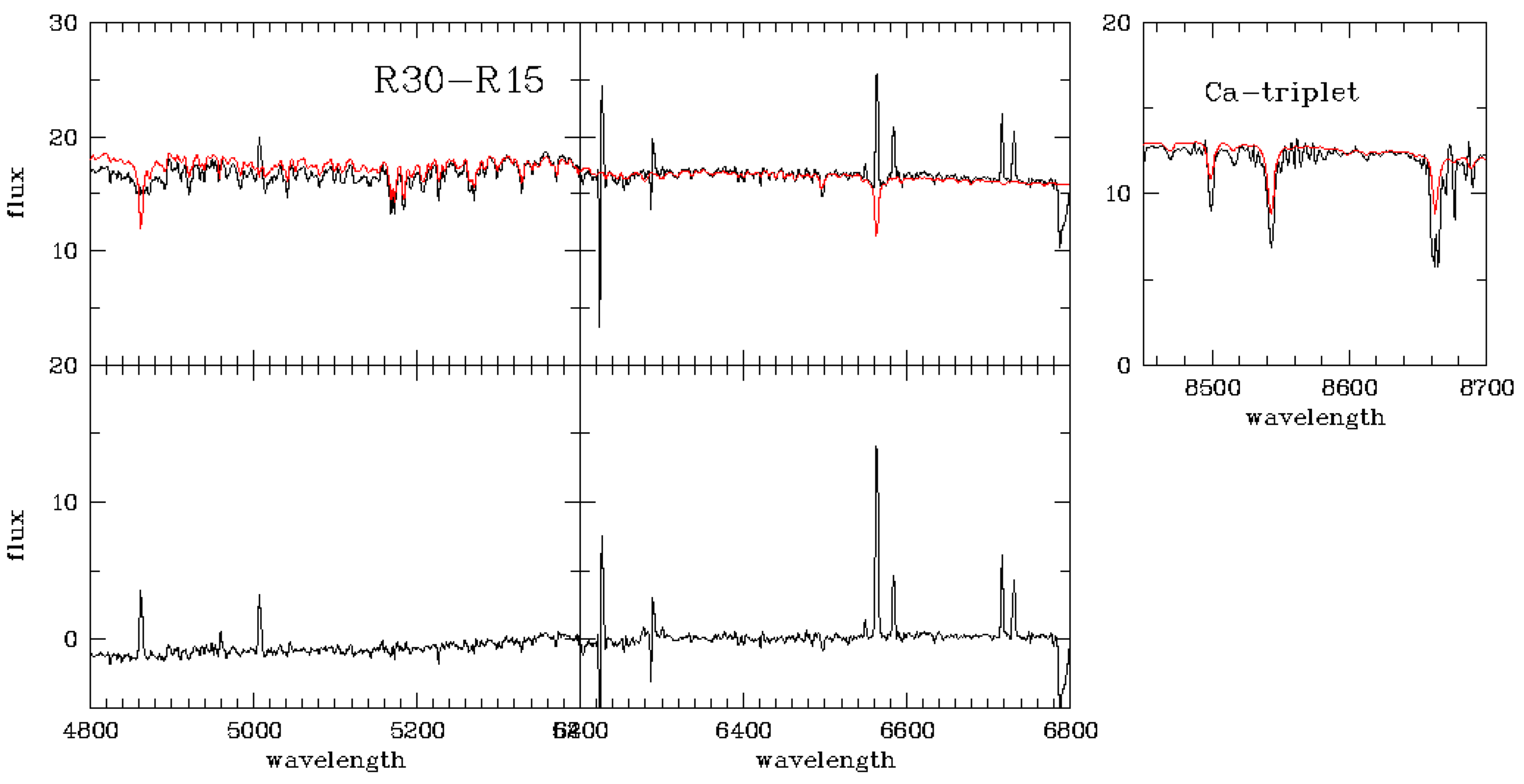}
\caption{The spectrum of R30-R15, which is the outer region of the dwarf.  The emission lines stem from the faint north-western parts of the gaseous ring.  
 The flux unit here is $\mathrm 10^{-17} erg/s/ \AA$.    }


\end{center}
\label{fig:regionB}
\end{figure*}

\end{appendix}

\end{document}